\documentclass[%
 reprint,
 amsmath,amssymb,
 aps,
]{revtex4-2}
\usepackage{placeins}
\usepackage{graphicx}
\usepackage{dcolumn}
\usepackage{bm}
\usepackage[T1]{fontenc}
\usepackage[utf8]{inputenc}
\usepackage{float}
\usepackage[colorlinks=true,linkcolor=blue,citecolor=blue]{hyperref}

\begin{document}

\title{Non-Markovian Decoherence Times in Finite-Memory Environments}

\author{Ramandeep Dewan}
\affiliation{Department of Physics, SVNIT, Surat}

\date{\today}

\begin{abstract}
Decoherence is often modeled using Markovian master equations that predict exponential suppression of coherence and are frequently used as effective bounds on quantum behavior in complex environments. Such descriptions, however, correspond to the singular physical limit of vanishing environmental memory. Here we formulate decoherence using a general time-nonlocal decoherence functional determined solely by the environmental force correlation function, with Markovian dynamics recovered explicitly as a limiting case.

For arbitrary stationary environments with finite temporal correlations, we show that the decoherence functional exhibits quadratic short-time growth that is model-independent within the finite-memory class considered. Consequently, the decoherence time defined operationally—without assuming exponential decay—scales as the square root of the environmental correlation time, independent of the detailed form of the bath correlation kernel.

These results are illustrated analytically for Gaussian-correlated, soft power-law, and Ornstein–Uhlenbeck environments. In the Ornstein–Uhlenbeck case, the non-Markovian dynamics admit an exact analytical closure, yielding a closed evolution equation for the coherence. Exact numerical simulations based on a pseudomode mapping confirm the predicted scaling and show that exponential decoherence emerges only in the memoryless limit.

Beyond coherence decay, we distinguish decoherence rates from observable loss of quantum signatures by analyzing purity and von Neumann entropy dynamics. We show that suppression of a specific coherence element need not coincide with irreversible entropy production. Finally, we introduce an inferred-memory perspective in which the environmental correlation time is treated as an operationally extractable parameter from dynamical data.
\end{abstract}

\maketitle

\section{Introduction}

Understanding how quantum coherence is suppressed by environmental interactions is a central problem in the theory of open quantum systems \cite{BreuerPetruccione2002,Weiss2012,Joos2003,Zurek2003}. 
Decoherence underlies the emergence of classical behavior, limits the performance of quantum technologies, and constrains the observability of quantum effects in complex environments \cite{Zurek1991,Zurek2003}. 
In many practical treatments, environmental noise is assumed to be memoryless, leading to Markovian master equations and exponential decay of coherence \cite{Lindblad1976,Gorini1976}. 
While mathematically convenient, this approximation corresponds to a singular physical limit that is rarely realized exactly in nature \cite{BreuerPetruccione2002,deVegaAlonso2017}.

A large body of work has established that non-Markovian environments-characterized by finite temporal correlations-can qualitatively alter reduced system dynamics \cite{Breuer2016,Rivas2014,deVegaAlonso2017}. 
Memory effects give rise to quadratic short-time behavior \cite{Khalfin1958,MisraSudarshan1977,Facchi2002}, 
information backflow \cite{Breuer2009,Lorenzo2013}, 
delayed entropy production \cite{Breuer2007,Rivas2010}, 
and deviations from simple exponential decay \cite{KofmanKurizki2000,KofmanKurizki2001}. 
These features are not restricted to specific microscopic models but reflect general structural properties of time-correlated environments \cite{Breuer2016}.

Despite this, Markovian decoherence laws continue to be widely employed as effective bounds, particularly in complex or strongly interacting settings where microscopic modeling is difficult \cite{JoosZeh,BreuerPetruccione2002}. 
A notable example is Tegmark’s widely cited estimate of decoherence times in biological media \cite{Tegmark2000}, 
which assumes delta-correlated environmental noise and yields an exponential suppression of spatial superpositions on extremely short timescales. 
While the calculation itself is internally consistent, its physical interpretation relies critically on the assumption of a strictly memoryless environment \cite{BreuerPetruccione2002}.

The central aim of this work is to place such Markovian decoherence bounds within a broader and more general theoretical framework. 
Rather than modifying specific models, we formulate decoherence in terms of a general time-nonlocal decoherence functional that depends only on the environmental force correlation function ~\cite{FeynmanVernon1963,caldeira1983,Grabert1988}. 
This approach allows us to identify the Markovian exponential decay law as a singular limit of vanishing environmental memory and to derive the generic behavior that arises for any finite-memory environment within the stated assumptions. Throughout this work, statements of generality are understood to apply to stationary environments with continuous correlation functions, linear system--bath coupling, and initially factorized system--environment states. Environments with strong initial correlations, nonstationary statistics, or singular noise spectra may exhibit qualitatively different short-time behavior and are not considered here.

We show that for arbitrary bath correlation kernels with finite temporal width, the coherence exhibits a quadratic short-time decay \cite{Khalfin1958,Facchi2002}. 
As a direct consequence, the decoherence time defined operationally-without assuming exponential decay-scales as the square root of the environmental correlation time. 
This result holds independently of the detailed shape of the correlation function and reflects a generic suppression of decoherence by environmental memory \cite{Breuer2016}.

To make these conclusions concrete, we analyze several representative finite-memory environments, including Gaussian, soft power-law, and Ornstein-Uhlenbeck correlation kernels \cite{GardinerZoller2004,UhlenbeckOrnstein1930}. 
The Ornstein-Uhlenbeck case is distinguished by the fact that it admits an exact analytical closure of the non-Markovian dynamics \cite{DiosiStrunz1998,Strunz1999}, 
allowing the coherence evolution to be obtained in closed form. 
Exact numerical simulations based on a pseudomode mapping confirm the analytical predictions \cite{Garraway1997} 
and demonstrate the emergence of exponential decay only in the singular Markovian limit.

Beyond coherence decay, we emphasize the distinction between decoherence rates and irreversible loss of quantum information \cite{Breuer2007}. 
In non-Markovian systems, coherence suppression, purity loss, and entropy production need not occur on the same timescale \cite{Rivas2010}. 
We analyze these quantities explicitly and show that finite environmental memory delays entropy growth even when coherence is already suppressed.

Although biological environments provide a motivating context for parts of this work, the results are not specific to biological systems. 
The finite-memory decoherence framework developed here applies equally to engineered quantum platforms such as superconducting qubits, cavity QED systems, and trapped ions subjected to colored noise \cite{Ithier2005,Bylander2011}. 
By treating the environmental correlation time as an inferred parameter rather than an assumed one, the framework offers a model-agnostic route for identifying non-Markovian signatures in experimental data \cite{Breuer2009}.  
We emphasize that finite environmental memory alone does not imply the presence of long-lived or functionally relevant quantum coherence, but rather modifies the structure of decoherence at short and intermediate times.

The novelty of the present work does not lie in the existence of quadratic short-time decoherence itself, which has appeared implicitly in several prior contexts, but in its systematic reinterpretation as an operational diagnostic of environmental memory. By treating the bath correlation time as an inferred parameter rather than an assumed microscopic constant, we place commonly used Markovian decoherence bounds within a unified finite-memory framework and clarify their status as singular limiting cases.

The paper is organized as follows. In Sec.~\ref{sec:decoh_func} we formulate decoherence in terms of a general time-nonlocal decoherence functional and identify the Markovian limit. 
In Sec.~\ref{sec:generic_short_time} we derive the generic quadratic short-time decoherence law for finite-memory environments within the stated assumptions. 
Sec.~\ref{sec:examples} presents explicit analytical results for Gaussian, soft power-law, and Ornstein-Uhlenbeck baths. 
Exact numerical validation using a pseudomode mapping is provided in Sec.~\ref{sec:numerics}. 
Secs.~\ref{sec:rate_vs_obs} and \ref{sec:inferred_tau} examine the distinction between decoherence and observable quantum signatures and introduce an inferred-memory perspective. 
Applications to clean physical systems and detailed derivations are presented in the appendices, followed by concluding remarks.

An earlier, biologically motivated analysis of finite-memory effects in decoherence was presented in Ref.~\cite{Dewan2026FiniteMemoryBio}. The present work substantially generalizes and reframes those results within a model-agnostic, operational framework, separating decoherence rates from observable quantum signatures and treating the environmental correlation time as an inferred parameter.

\section{Decoherence Functional and the Markovian Limit}
\label{sec:decoh_func}

We consider a generic open quantum system coupled linearly to an environmental bath \cite{BreuerPetruccione2002,Weiss2012}. The total Hamiltonian is taken to be
\begin{equation}
H = H_S + H_E + x \otimes F ,
\end{equation}
where $H_S$ is the system Hamiltonian, $H_E$ describes the environment, $x$ is a system operator distinguishing two localized states separated by a distance $a$, and $F$ is an environmental force operator. The influence of the environment on the system is fully characterized by the bath correlation function
\begin{equation}
\alpha(t,s) = \langle F(t) F(s) \rangle ,
\end{equation}
where the expectation value is taken with respect to the initial bath state \cite{BreuerPetruccione2002,PazZurek1993}.

We assume an initially factorized state $\rho_{\mathrm{tot}}(0)=\rho(0)\otimes\rho_E$ and focus on the decay of the off-diagonal coherence element \cite{BreuerPetruccione2002,Grabert1988}
\begin{equation}
C(t) = \langle L | \rho(t) | R \rangle ,
\end{equation}
where $|L\rangle$ and $|R\rangle$ are eigenstates of $x$ corresponding to two spatially separated configurations.

For linear system-bath coupling, the reduced coherence can be expressed in terms of a decoherence functional $\Phi(t)$ as
\begin{equation}
C(t) = C(0)\, e^{-\Phi(t)} ,
\end{equation}
where the decoherence functional is given by
\begin{equation}
\Phi(t) = \frac{a^2}{\hbar^2}
\int_0^t ds \int_0^t ds'\, \alpha(s,s') .
\label{eq:decoh_functional}
\end{equation}
Equation~\eqref{eq:decoh_functional} assumes an initially factorized system-environment state, linear system-bath coupling, and stationary bath correlations \cite{FeynmanVernon1963,caldeira1983,Grabert1988}. Initial system--environment correlations or explicitly nonstationary environments can modify the short-time structure of decoherence, including the possible appearance of linear-in-time decay even in non-Markovian settings. Such effects lie outside the scope of the present framework and are not considered here.
Equation~\eqref{eq:decoh_functional} provides a fully non-Markovian expression for coherence decay. A detailed derivation and its generic short-time expansion are presented in Appendix~\ref{app:generic}.
Equation~\eqref{eq:decoh_functional} is general for systems with linear coupling to environments admitting an influence-functional description, including Gaussian quantum baths and classical stochastic environments with identical correlation functions. It does not assume Markovianity, but it does rely on the absence of initial system-environment correlations and on stationary bath statistics.

\subsection{Markovian Limit}

The Markovian approximation corresponds to the assumption that environmental correlations decay instantaneously compared to system timescales \cite{Lindblad1976,Gorini1976}. This is represented by a delta-correlated bath,
\begin{equation}
\alpha(s,s') = 2D\,\delta(s-s') ,
\label{eq:markov_alpha}
\end{equation}
where $D$ is the noise strength determined by microscopic collision processes \cite{JoosZeh,Zurek2003}.

Substituting Eq.~\eqref{eq:markov_alpha} into the decoherence functional~\eqref{eq:decoh_functional} yields
\begin{equation}
\Phi(t) = \frac{2 a^2 D}{\hbar^2} \int_0^t ds = \frac{a^2 D}{\hbar^2}\, t .
\end{equation}
The coherence therefore decays exponentially,
\begin{equation}
C(t) = C(0)\exp\!\left(-\frac{a^2 D}{\hbar^2} t\right),
\end{equation}
with a characteristic decoherence time
\begin{equation}
\tau_{\mathrm{M}} = \frac{\hbar^2}{a^2 D}.
\end{equation}
This Markovian exponential decay law is recovered within the present framework as the singular limit of vanishing bath correlation time; its relation to commonly used Markovian decoherence estimates in biological media is discussed in Appendix~\ref{app:tegmark} \cite{Tegmark2000}.
This exponential decay law is widely employed in decoherence estimates and underlies many previously proposed bounds on coherence lifetimes \cite{JoosZeh,Zurek2003}. However, it follows directly from the singular assumption~\eqref{eq:markov_alpha} that the bath correlation time vanishes identically. As we show in the following sections, this limit qualitatively alters the short-time structure of the decoherence functional.

\subsection{Finite-Memory Environments}

For environments with finite temporal correlations, the bath correlation function $\alpha(s,s')$ has a characteristic width $\tau_c$ \cite{Kubo,vanKampen}. In this case the double integral in Eq.~\eqref{eq:decoh_functional} no longer reduces to a linear function of time. Instead, the short-time behavior of $\Phi(t)$ is governed by the continuity of $\alpha(s,s')$ near $s=s'$. Throughout this work we therefore restrict attention to initially uncorrelated system-environment states; the effects of initial correlations on short-time decoherence are well known but are not considered here.

As a result, finite-memory environments generically produce decoherence dynamics that differ qualitatively from the Markovian exponential law at early times \cite{deVegaAlonso2017}. In the next section, we show that for any bath with finite correlation time, the decoherence functional grows quadratically at short times, leading to a decoherence time that scales as the square root of the bath correlation time. The familiar Markovian result emerges only as a singular limiting case. The present analysis assumes stationary bath correlations; extensions to driven or explicitly non-equilibrium environments, where time-translation invariance is broken, remain an important direction for future work.

\section{Generic Short-Time Decoherence for Finite-Memory Baths}
\label{sec:generic_short_time}

We now analyze the short-time behavior of the decoherence functional~\eqref{eq:decoh_functional} for environments with finite temporal correlations \cite{Khalfin1958,Facchi2002}. Our goal is to establish results that are independent of the detailed microscopic structure of the bath and depend only on the existence of a finite correlation time.

We consider bath correlation functions $\alpha(s,s')$ that are continuous and sufficiently well-behaved near $s=s'$, with a characteristic correlation time $\tau_c$ beyond which correlations decay appreciably \cite{BreuerPetruccione2002,deVegaAlonso2017}. This assumption encompasses a broad class of physical environments, including Gaussian-correlated noise, exponentially correlated (Ornstein-Uhlenbeck) noise, and baths with soft power-law cutoffs \cite{Kubo,vanKampen}.

\subsection{Short-Time Expansion of the Decoherence Functional}

For times $t$ short compared to the bath correlation time, $t \ll \tau_c$, the bath correlation function varies slowly over the integration domain \cite{MisraSudarshan1977,Facchi2001}. In this regime, $\alpha(s,s')$ may be expanded about $s=s'$ as
\begin{equation}
\alpha(s,s') = \alpha(0,0) + O(|s-s'|/\tau_c) .
\end{equation}
Retaining only the leading contribution, the decoherence functional~\eqref{eq:decoh_functional} becomes
\begin{equation}
\Phi(t) \approx \frac{a^2}{\hbar^2}\,\alpha(0,0)
\int_0^t ds \int_0^t ds' .
\end{equation}
The double integral evaluates straightforwardly, yielding
\begin{equation}
\Phi(t) \approx \frac{a^2}{\hbar^2}\,\alpha(0,0)\, t^2 .
\label{eq:quadratic_phi}
\end{equation}
Thus, for any bath with finite temporal correlations, the decoherence functional grows quadratically at short times \cite{Khalfin1958,Facchi2002}. We emphasize that this generic quadratic behavior assumes stationary bath correlations that are continuous at coincident times and an initially factorized system-environment state. Environments with singular short-time correlations, nonstationary noise, or strong initial system-bath correlations may exhibit modified short-time structure and fall outside the scope of the present analysis. As a consequence, the coherence obeys
\begin{equation}
C(t) \approx C(0)\left(1 - \frac{a^2}{\hbar^2}\,\alpha(0,0)\, t^2 \right),
\end{equation}
demonstrating that quadratic short-time decoherence is a generic feature of finite-memory environments within the stated assumptions.

This behavior contrasts sharply with the linear-in-time growth of $\Phi(t)$ obtained under the Markovian approximation, which relies on the singular assumption of a delta-correlated bath \cite{BreuerPetruccione2002}. The quadratic decay reflects the continuity of environmental fluctuations and is consistent with general analyses of short-time dynamics in non-Markovian open quantum systems \cite{deVegaAlonso2017,RivasHuelgaPlenio}.

\subsection{Decoherence Time and Scaling with Bath Memory}

To characterize the timescale of coherence loss, we define the decoherence time $\tau_{\mathrm{dec}}$ operationally via
\begin{equation}
\Phi(\tau_{\mathrm{dec}}) = 1 ,
\end{equation}
corresponding to a reduction of coherence by a factor $e^{-1}$.

Using the quadratic short-time form~\eqref{eq:quadratic_phi}, we obtain
\begin{equation}
\tau_{\mathrm{dec}} \sim
\sqrt{\frac{\hbar^2}{a^2\,\alpha(0,0)}} .
\label{eq:tau_dec_general}
\end{equation}
The bath-dependent quantity $\alpha(0,0)$ may be related to the bath correlation time by noting that, for a wide class of finite-memory environments, the zero-time correlation scales as
\begin{equation}
\alpha(0,0) \sim \frac{D}{\tau_c},
\end{equation}
where $D$ sets the overall noise strength \cite{Kubo,vanKampen}. This scaling reflects the normalization of finite-memory kernels, for which the integrated noise strength $\int d\tau\,\alpha(\tau)=D$ is held fixed as $\tau_c$ varies. Physically, this corresponds to fixing the long-time Markovian noise strength while varying the memory time, ensuring a meaningful comparison between environments with different correlation widths. Substituting this scaling into Eq.~\eqref{eq:tau_dec_general} yields
\begin{equation}
\tau_{\mathrm{dec}} \sim \sqrt{\frac{\hbar^2\,\tau_c}{a^2 D}} .
\label{eq:sqrt_scaling}
\end{equation}
The square-root scaling characterizes the onset of decoherence for times short compared to the bath correlation time; higher-order corrections become relevant only at times $t \sim \tau_c$, beyond which the detailed kernel shape influences the dynamics.
The generality of this scaling, including its independence from the detailed form of the bath correlation kernel, is demonstrated explicitly in Appendix~\ref{app:generic}.

Equation~\eqref{eq:sqrt_scaling} constitutes the central result of this section. It shows that for any environment with finite correlation time, the decoherence time scales as the square root of the bath memory timescale. This scaling is independent of the detailed form of the bath correlation function and depends only on the existence of a finite temporal width.

Quadratic short-time decay of survival probabilities and coherence has appeared previously in contexts such as the Khalfin theorem and quantum Zeno dynamics. In contrast to those works, which emphasize measurement-induced effects or spectral constraints, the present analysis isolates finite environmental memory as the sole structural origin of quadratic decoherence within a unified influence-functional framework.

\subsection{Singular Nature of the Markovian Limit}

The Markovian exponential decay law is recovered formally by taking the limit $\tau_c \to 0$ while keeping $D$ fixed \cite{Lindblad1976,Gorini1976}. In this limit, $\alpha(0,0)$ diverges and the quadratic short-time regime collapses, yielding a decoherence functional linear in time. This demonstrates that the Markovian result corresponds to a singular limit in which environmental memory is removed entirely.

The finite-memory scaling~\eqref{eq:sqrt_scaling} therefore represents the generic behavior of decoherence in realistic environments within the stated assumptions, while the familiar Markovian decoherence time emerges only as a special limiting case.

\section{Representative Finite-Memory Environments}
\label{sec:examples}

To illustrate the general finite-memory decoherence framework developed in the previous sections, we now analyze several representative bath correlation kernels \cite{Kubo,vanKampen}. These examples are chosen to span a broad class of physically relevant environments while remaining analytically tractable. In each case, we demonstrate explicitly how the quadratic short-time decoherence law and the square-root scaling of the decoherence time emerge. Detailed derivations are deferred to the appendices.

\subsection{Gaussian-Correlated Environment}

We first consider a bath with Gaussian temporal correlations,
\begin{equation}
\alpha(t,s) = \frac{D}{\sqrt{\pi}\tau_c}\exp\!\left[-\frac{(t-s)^2}{\tau_c^2}\right],
\label{eq:gaussian_kernel}
\end{equation}
where $D$ sets the noise strength and $\tau_c$ characterizes the width of the correlation function. Such kernels arise naturally in environments with smooth spectral cutoffs and finite-bandwidth noise \cite{caldeira1983,Grabert1988}.

Substituting Eq.~\eqref{eq:gaussian_kernel} into the general decoherence functional~\eqref{eq:decoh_functional} and expanding for times $t \ll \tau_c$, one finds
\begin{equation}
\Phi(t) \approx \frac{a^2 D}{\hbar^2 \tau_c} t^2 + O(t^4),
\end{equation}
consistent with the generic quadratic short-time behavior within the stated assumptions derived in Sec.~\ref{sec:generic_short_time} \cite{Khalfin1958,Facchi2002}. The corresponding decoherence time therefore scales as
\begin{equation}
\tau_{\mathrm{dec}} \sim \sqrt{\frac{\hbar^2 \tau_c}{a^2 D}} .
\end{equation}

The full time dependence of $\Phi(t)$ involves error functions and does not admit a simple closed-form exponential representation \cite{Grabert1988}. However, the short-time scaling and the singular recovery of Markovian behavior in the limit $\tau_c \to 0$ are transparent. The full derivation of the Gaussian-kernel decoherence functional is provided in Appendix~\ref{app:generic}.

\subsection{Soft Power-Law Correlated Environment}

Next, we consider a bath with soft power-law temporal correlations,
\begin{equation}
\alpha(t,s) = \frac{D}{\tau_c}\frac{1}{\left[1 + \left(\frac{|t-s|}{\tau_c}\right)^p\right]},
\label{eq:powerlaw_kernel}
\end{equation}
with exponent $p>1$ ensuring integrability. Such kernels model environments with long-tailed correlations arising from structured or hierarchical baths and appear in systems with soft spectral cutoffs \cite{Weiss2012,KofmanKurizki2000}.

For finite $\tau_c$, the correlation function remains continuous at $t=s$, and the decoherence functional may again be expanded for short times. Retaining the leading contribution yields
\begin{equation}
\Phi(t) \approx \frac{a^2 D}{\hbar^2 \tau_c} t^2 + O(t^{2+p}),
\end{equation}
independent of the specific value of $p$.

The resulting decoherence time obeys the same square-root scaling,
\begin{equation}
\tau_{\mathrm{dec}} \sim \sqrt{\frac{\hbar^2 \tau_c}{a^2 D}},
\end{equation}
demonstrating that the finite-memory enhancement of coherence lifetimes is robust even in the presence of slowly decaying correlations \cite{KofmanKurizki2001}. The Markovian exponential decay law is recovered only in the singular limit $\tau_c \to 0$, where the power-law kernel collapses to a delta function. Details of the short-time expansion for soft power-law correlations are given in Appendix~\ref{app:generic}.

\subsection{Ornstein-Uhlenbeck Environment}

Finally, we consider the Ornstein-Uhlenbeck (OU) environment, characterized by an exponentially decaying correlation function,
\begin{equation}
\alpha(t,s) = \frac{D}{\tau_c} e^{-|t-s|/\tau_c}.
\label{eq:ou_kernel}
\end{equation}
The OU bath provides a minimal model of colored noise with finite correlation time and plays a distinguished role due to its exact solvability \cite{UhlenbeckOrnstein1930}.

For this kernel, the non-Markovian dynamics of the coherence can be solved analytically using the non-Markovian quantum state diffusion formalism \cite{Diosi1998,Strunz1999}. The functional derivative terms close exactly for exponential correlations, leading to a second-order differential equation for $C(t)$,
\begin{equation}
\ddot{C}(t) + \frac{1}{\tau_c}\dot{C}(t) + \frac{2a^2 D}{\hbar^2 \tau_c} C(t) = 0.
\label{eq:ou_ode}
\end{equation}
The solution is uniquely specified by the initial conditions
$C(0)=1$ and $\dot{C}(0)=0$, corresponding to an initially pure superposition state.
The solution of Eq.~\eqref{eq:ou_ode} yields the full time dependence of the coherence and explicitly realizes the quadratic short-time decay,
\begin{equation}
C(t) \approx 1 - \frac{a^2 D}{\hbar^2 \tau_c} t^2 + O(t^3).
\end{equation}

Defining the decoherence time via $C(\tau_{\mathrm{dec}})=e^{-1}$, one obtains
\begin{equation}
\tau_{\mathrm{dec}} = \sqrt{\frac{\hbar^2 \tau_c}{a^2 D}},
\end{equation}
in agreement with the general finite-memory scaling. In the singular limit $\tau_c \to 0$, the OU kernel reduces to a delta-correlated bath and the solution of Eq.~\eqref{eq:ou_ode} converges to the familiar Markovian exponential decay law \cite{BreuerPetruccione2002}.

The exact non-Markovian closure leading to Eq.~\eqref{eq:ou_ode} is derived in Appendix~\ref{app:ou_closure}. 
The Ornstein-Uhlenbeck kernel is distinguished among the examples considered in that its exponential form permits an exact analytic closure of the non-Markovian dynamics; Gaussian and power-law kernels do not admit such a closed evolution equation, although their short-time behavior is fully captured by the general framework.

\section{Exact Numerical Validation}
\label{sec:numerics}

To validate the analytical finite-memory decoherence scaling derived in the previous sections, we perform exact numerical simulations for environments with exponential correlations using the pseudomode mapping \cite{Garraway1997a,Garraway1997b}. This approach provides a non-perturbative and numerically exact representation of non-Markovian dynamics for baths with single-exponential correlation functions. While numerical validation is presented here for exponentially correlated environments, the analytical finite-memory scaling does not rely on this specific kernel and applies more generally to baths with finite correlation time.

\subsection{Pseudomode Representation}

For the Ornstein-Uhlenbeck environment, the bath correlation function
\begin{equation}
\alpha(t,s) = \frac{D}{\tau_c} e^{-|t-s|/\tau_c}
\end{equation}
can be represented exactly by coupling the system to an auxiliary damped harmonic mode (the pseudomode), which in turn is Markovianly coupled to a memoryless reservoir \cite{Garraway1997a}. This mapping transforms the original non-Markovian problem into a Markovian Lindblad evolution in an enlarged Hilbert space and has been shown to be exact for exponential bath correlations \cite{Imamoglu1994,Garraway1997b}.

We numerically integrate the resulting Lindblad master equation and extract the coherence $C(t)$ for a range of bath correlation times $\tau_c$. The decoherence time $\tau_{\mathrm{dec}}$ is defined operationally by the condition $|C(\tau_{\mathrm{dec}})| = e^{-1} |C(0)|$.

\subsection{Decoherence Time Scaling}

\begin{figure}[H]
    \centering
    \includegraphics[width=0.48\textwidth]{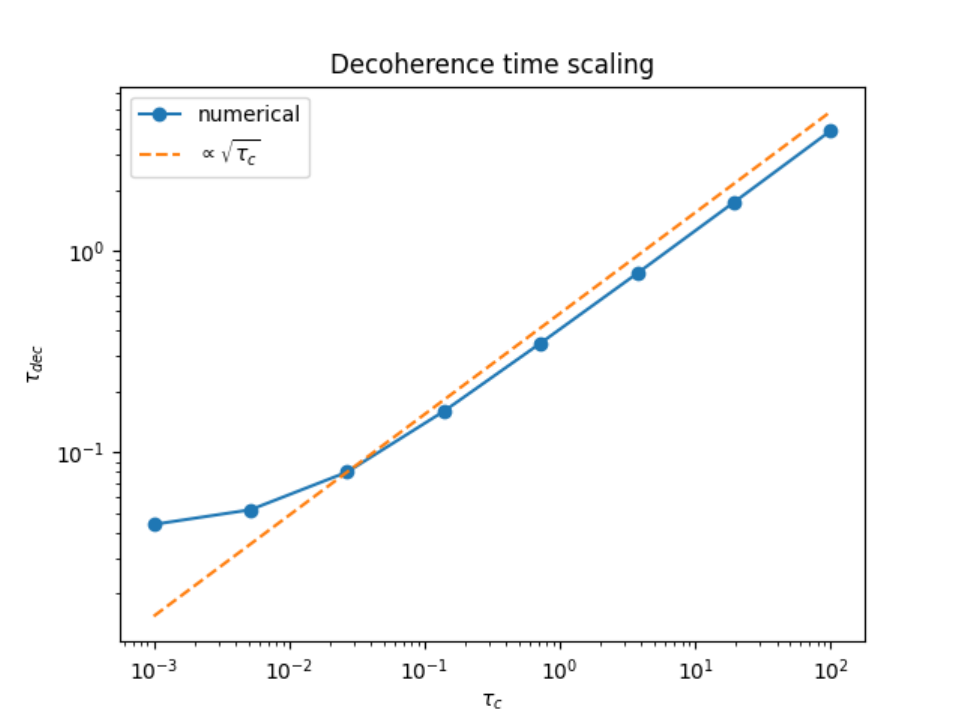}
    \caption{Decoherence time $\tau_{\mathrm{dec}}$ extracted from pseudomode simulations as a function of bath correlation time $\tau_c$. The dashed line indicates the $\sqrt{\tau_c}$ scaling predicted analytically.(Data and simulation parameters as in Ref.~\cite{Dewan2026FiniteMemoryBio}; shown here to illustrate the generalized finite-memory scaling discussed in the present work.)}
    \label{fig:scaling}
\end{figure}

Figure~\ref{fig:scaling} shows the decoherence time $\tau_{\mathrm{dec}}$ obtained from the pseudomode simulations as a function of the bath correlation time $\tau_c$. The numerical data exhibit a clear square-root dependence,
\begin{equation}
\tau_{\mathrm{dec}} \propto \sqrt{\tau_c},
\end{equation}
in excellent agreement with the analytical prediction derived from the finite-memory decoherence functional.
Representative coherence decay curves and numerical details of the pseudomode simulations are provided in Appendix~\ref{app:numerics}.
The numerical results also confirm the collapse of the quadratic short-time regime in the singular limit $\tau_c \to 0$, where the coherence dynamics approaches the Markovian exponential decay law \cite{BreuerPetruccione2002}.
We verified that the extracted scaling is insensitive to moderate variations in numerical resolution, simulation time window, and pseudomode damping parameters.

\subsection{Consistency with Hierarchical Equation Methods}

For environments with exponential correlations, the pseudomode mapping is formally equivalent to the lowest-tier truncation of the hierarchical equations of motion (HEOM) \cite{TanimuraKubo1989,IshizakiTanimura2005}. Both approaches provide exact representations of the same underlying non-Markovian dynamics. As a result, convergence with respect to hierarchy depth is guaranteed for the Ornstein-Uhlenbeck bath, and independent HEOM simulations would reproduce the same coherence dynamics obtained via the pseudomode method. The equivalence between the pseudomode mapping and the lowest-tier truncation of the hierarchical equations of motion is discussed in Appendix~\ref{app:numerics}.

The present numerical results therefore provide an exact non-perturbative confirmation of the analytical finite-memory decoherence scaling for exponentially correlated environments. The Ornstein-Uhlenbeck environment serves here as a representative exactly solvable finite-memory model for numerical validation. The emergence of quadratic short-time decay and square-root decoherence-time scaling in this case supports the analytical prediction that these features depend only on finite temporal correlations, rather than on the detailed functional form of the bath kernel.

\section{Decoherence Rate versus Observable Quantum Signatures}
\label{sec:rate_vs_obs}

In Markovian open quantum systems, decoherence is commonly characterized by a single exponential decay rate, and the associated decoherence time is often implicitly identified with the loss of all observable quantum features \cite{BreuerPetruccione2002,Lindblad1976}. This identification underlies many classicality arguments based on exponential decoherence laws, including Tegmark’s original analysis \cite{Tegmark2000}. However, this correspondence is no longer valid in the presence of environmental memory \cite{Breuer2009,Rivas2010}. Explicit expressions for purity and von Neumann entropy dynamics corresponding to the finite-memory decoherence law are derived in Appendix~\ref{app:purity_entropy}. Here we use the term classicalization to denote irreversible entropy production and loss of recoverable quantum information, rather than the suppression of a single coherence element.

For finite-memory environments, decoherence is intrinsically non-Markovian and cannot be fully characterized by a single decay constant \cite{Breuer2016}. As established in Sec.~\ref{sec:generic_short_time}, finite-memory environments induce a separation of timescales between coherence decay and entropy production. In particular, the decay of off-diagonal coherence elements, the loss of state purity, and the growth of von Neumann entropy are no longer synchronized processes \cite{Zyczkowski1998,Breuer2007}.

To illustrate this distinction, consider three commonly used diagnostics of open-system dynamics \cite{NielsenChuang}:  
\begin{itemize}
    \item the coherence $C(t)$,  
    \item the purity $\mathcal{P}(t) = \mathrm{Tr}[\rho(t)^2]$, and  
    \item the von Neumann entropy $S(t) = -\mathrm{Tr}[\rho(t)\ln\rho(t)]$.  
\end{itemize}
Within this framework, coherence decay provides the most direct diagnostic of environmental memory, while entropy production is more closely associated with irreversible classicalization. Purity loss typically interpolates between these behaviors and may exhibit delayed or nonmonotonic dynamics in non-Markovian regimes.
In Markovian dynamics with exponential decoherence, all three quantities exhibit monotonic behavior governed by the same characteristic timescale~\cite{Lindblad1976}. In contrast, finite-memory environments generically give rise to delayed decoherence, partial revivals, and extended plateaus in these quantities~\cite{Breuer2009,Lorenzo2013}. As a result, a short decoherence time extracted from the decay of $C(t)$ does not necessarily imply rapid entropy production or irreversible classicalization~\cite{Breuer2007,Rivas2010}.

This separation between the decoherence rate and the observable loss of quantum signatures is a well-known feature of non-Markovian dynamics \cite{Breuer2016,deVegaAlonso2017}, but it is often obscured when Markovian approximations are implicitly assumed. In the present framework, this separation emerges naturally as a consequence of finite bath correlation time and does not rely on any system-specific assumptions.

Importantly, the existence of a finite decoherence time does not imply the complete suppression of quantum effects. Instead, it characterizes the timescale over which a particular coherence measure decays under a chosen operational definition. In non-Markovian environments, quantum features may persist beyond this timescale in the form of reduced entropy production, coherence revivals, or delayed equilibration \cite{Breuer2009,Rivas2010}.

The results of this work therefore clarify that decoherence times extracted from Markovian approximations should not be interpreted as definitive indicators of classical behavior when environmental memory is present. Any assessment of observable quantum effects in structured environments must distinguish between the rate of coherence decay and the broader dynamical signatures of non-Markovian evolution. Here we use classicalization to denote irreversible entropy production and loss of recoverable quantum information, rather than a mere suppression of a specific coherence element \cite{Zurek2003}.

\section{Environmental Memory as an Inferred Parameter}
\label{sec:inferred_tau}

A central advantage of the finite-memory decoherence framework developed in this work is that it avoids the need to assume microscopic environmental properties \emph{a priori} \cite{BreuerPetruccione2002,deVegaAlonso2017}. In particular, the bath correlation time $\tau_c$ enters the decoherence functional as a control parameter governing the crossover between Markovian and non-Markovian dynamical regimes~\cite{Breuer2016,Rivas2014}. A detailed formulation of the inferred correlation-time framework and its operational interpretation are presented in Appendix~\ref{app:inferred_tau}.

The analytical and numerical results of Secs.~\ref{sec:generic_short_time}-\ref{sec:numerics} demonstrate that finite bath memory manifests through a model-independent within the finite-memory class considered quadratic short-time coherence decay and a corresponding $\tau_{\mathrm{dec}} \propto \sqrt{\tau_c}$ scaling. These features provide the basis for operational inference of environmental memory effects.

The analytical results derived in Secs.~\ref{sec:generic_short_time} and ~\ref{sec:examples} show that finite bath memory generically induces a quadratic short-time decay of coherence, with the associated decoherence time scaling as $\tau_{\mathrm{dec}} \propto \sqrt{\tau_c}$ \cite{KofmanKurizki2000,Facchi2002}. Numerical simulations in Sec.~\ref{sec:numerics} confirm this scaling and explicitly demonstrate the emergence of distinct dynamical regimes as $\tau_c$ is varied. For sufficiently small $\tau_c$, the dynamics rapidly approaches the Markovian exponential decay law, while larger values of $\tau_c$ give rise to delayed decoherence and enhanced persistence of coherence \cite{Breuer2009}.

Representative coherence decay curves illustrating this crossover between Markovian and non-Markovian regimes for different effective correlation times $\tau_c$ are shown in Appendix~\ref{app:numerics}. These results demonstrate how features such as delayed onset of decoherence and extended quadratic short-time behavior provide direct operational signatures of finite environmental memory \cite{Rivas2010}. Operational inference of $\tau_c$ from short-time curvature requires experimental time resolution sufficient to resolve times $t \ll \tau_c$; coarse-grained measurements may obscure the quadratic regime entirely.

This perspective shifts the focus from speculative questions about the quantum nature of specific environments to experimentally accessible criteria based on dynamical signatures \cite{Breuer2016}. By analyzing coherence decay curves and identifying deviations from Markovian exponential behavior, one can infer the effective bath correlation time without detailed microscopic modeling \cite{KofmanKurizki2001}. In this sense, the finite-memory decoherence functional provides a general diagnostic tool for environmental memory effects in open quantum systems, independent of the physical origin of the bath. Operational inference of the bath correlation time from coherence curvature requires experimental time resolution sufficient to resolve the regime $t \ll \tau_c$. Finite temporal resolution, statistical noise, or coarse-grained measurements may obscure the quadratic short-time behavior and lead to effective Markovian fits even in finite-memory environments. These limitations reflect experimental constraints rather than a breakdown of the finite-memory decoherence framework~\cite{Facchi2002}.

\section{Application to Controlled Open Quantum Systems}

Although the finite-memory decoherence framework developed in this work is formulated independently of any specific physical platform, it is particularly well suited to controlled open quantum systems in which environmental correlations can be engineered and probed directly \cite{BreuerPetruccione2002,deVegaAlonso2017}. In such systems, the bath correlation time $\tau_c$ is not an unknown microscopic property but an experimentally tunable parameter, providing a clean setting for testing non-Markovian decoherence laws \cite{Bylander2011,Murch2013}.

Examples include superconducting qubits coupled to structured electromagnetic environments \cite{Ithier2005,Bylander2011}, cavity and circuit QED systems with engineered reservoirs \cite{HarocheRaimond,Murch2013}, and trapped ions subjected to colored noise \cite{Myatt2000,Barreiro2011}. In these platforms, deviations from Markovian exponential decoherence have already been observed in related contexts, and the presence of finite bath memory can be controlled through circuit design, spectral filtering, or externally applied noise \cite{Bylander2011,Barreiro2011}.

Within the present framework, finite environmental memory leads to two generic and experimentally accessible signatures: a quadratic short-time suppression of coherence and a decoherence time that scales as $\tau_{\mathrm{dec}} \propto \sqrt{\tau_c}$ \cite{KofmanKurizki2000,Facchi2002}. These features provide clear diagnostics for identifying non-Markovian dynamics and distinguishing them from Markovian exponential decay \cite{Breuer2016}. Importantly, the predicted behavior does not depend on the microscopic origin of the bath, but only on the existence of finite temporal correlations.

Representative numerical illustrations of coherence dynamics for parameter regimes relevant to engineered open quantum systems are presented in Appendix~\ref{app:clean_system}. These results demonstrate how the finite-memory decoherence functional can be applied directly to clean physical systems without invoking system-specific assumptions or biological interpretations. As such, the framework developed here provides a general theoretical tool for analyzing decoherence in structured environments across a broad range of experimental platforms \cite{Rivas2014}. Numerical illustrations and model details for superconducting qubits subjected to colored noise are provided in Appendix~\ref{app:clean_system}.

\section{Conclusion}

We have developed a general finite-memory decoherence framework that clarifies the structure of decoherence in environments with temporal correlations. Starting from a time-nonlocal decoherence functional, we showed that quadratic short-time suppression of coherence is a generic consequence of finite bath memory within the stated assumptions, while the familiar exponential decay law arises only as a singular limiting case corresponding to vanishing correlation time \cite{BreuerPetruccione2002,Breuer2016}.

Explicit analytical results were obtained for several representative bath correlation models, including exponentially correlated, Gaussian, and soft power-law environments. For these cases, the decoherence time was shown to scale as the square root of the bath correlation time, in contrast to the inverse-noise-strength scaling predicted by Markovian theories. Exact numerical simulations based on a pseudomode mapping confirmed the analytical predictions for exponentially correlated environments. We emphasize that the present framework does not address environments with strong initial correlations, nonstationary noise, or singular spectral features, for which the short-time structure of decoherence may differ qualitatively.

Beyond the derivation of decoherence laws, we emphasized two conceptual points. First, in non-Markovian dynamics the decoherence rate should not be directly identified with the observable loss of quantum coherence, as different diagnostics such as coherence, purity, and entropy evolve on distinct timescales \cite{Breuer2007,Rivas2010}. Second, the bath correlation time should be viewed as an effective parameter that can be inferred from dynamical signatures rather than assumed from microscopic considerations.

The framework developed here applies broadly to structured environments in open quantum systems and provides a unified perspective on the role of environmental memory in decoherence processes. Applications to specific physical systems and biological contexts, including the recovery of previously proposed Markovian bounds as singular limits, are discussed in the appendices \cite{Tegmark2000}. More generally, the present results highlight the need to treat environmental memory explicitly when assessing decoherence times and the persistence of quantum coherence in realistic settings.
\paragraph{Limitations.} The finite-memory decoherence framework applies to initially factorized states with stationary correlations and linear coupling; extensions to initially correlated, strongly driven, or nonstationary environments may modify short-time behavior and require separate analysis.

\FloatBarrier
\appendix

\section{Numerical Coherence Dynamics}
\label{app:numerics}

This appendix provides numerical details supporting the exact simulations presented in Sec.~\ref{sec:numerics}. The results validate the analytical finite-memory decoherence scaling derived in the main text and illustrate the generic quadratic short-time behaviour of coherence for environments with finite correlation time within the stated assumptions.

\subsection{Representative Coherence Decay Curves}
\label{app:decay_curves}

\begin{figure}
    \centering
    \includegraphics[width=0.48\textwidth]{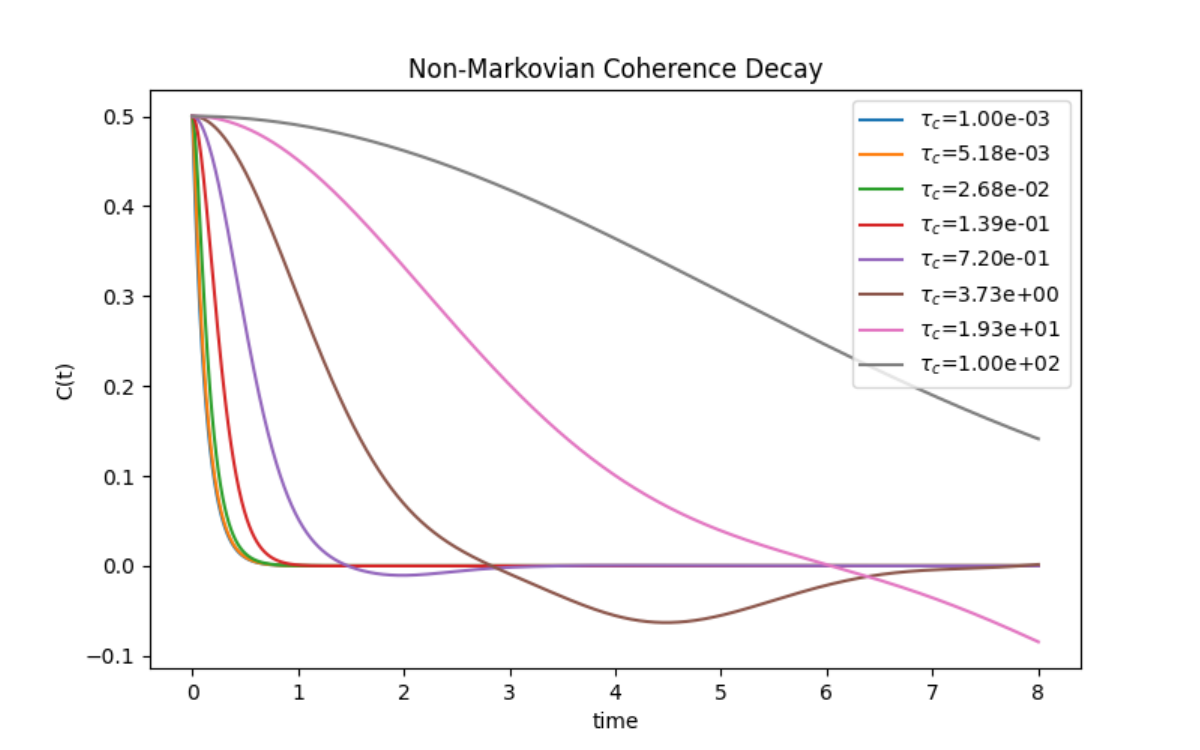}
    \caption{Representative coherence decay curves $C(t)$ obtained from exact pseudomode simulations for different bath correlation times $\tau_c$. For all finite $\tau_c$, the coherence exhibits a quadratic short-time decay, in contrast with the exponential behaviour predicted by the Markovian approximation.(Data and simulation parameters as in Ref.~\cite{Dewan2026FiniteMemoryBio}; shown here to illustrate the generalized finite-memory scaling discussed in the present work.)}
    \label{fig:decay}
\end{figure}

Figure~\ref{fig:decay} shows representative coherence decay curves $C(t)$ obtained from the pseudomode simulations for several values of the bath correlation time $\tau_c$. For finite $\tau_c$, the decay is clearly quadratic at short times, consistent with the general finite-memory decoherence functional derived in Sec.~\ref{sec:generic_short_time}. As $\tau_c$ is decreased, the quadratic regime progressively shrinks and the dynamics approaches the Markovian exponential decay law in the singular limit $\tau_c \to 0$.

\subsection{Extraction of the Decoherence Time}
\label{app:tau_dec}

The decoherence time $\tau_{\mathrm{dec}}$ is defined operationally by the condition
\begin{equation}
|C(\tau_{\mathrm{dec}})| = e^{-1} |C(0)| .
\end{equation}
For each bath correlation time $\tau_c$, the coherence $C(t)$ is sampled on a uniform time grid and the first time point satisfying the above condition is recorded as $\tau_{\mathrm{dec}}$. This definition is independent of any assumed functional form of the decay and is therefore applicable in both Markovian and non-Markovian regimes.

\subsection{Numerical Implementation}
\label{app:numerical_method}

The Ornstein-Uhlenbeck environment is simulated exactly using the pseudomode mapping, in which the structured bath is represented by a single auxiliary harmonic mode coupled to the system and damped by a memoryless reservoir~\cite{Garraway1997a,Garraway1997b}. This mapping transforms the original non-Markovian problem into a Markovian Lindblad evolution in an enlarged Hilbert space and is exact for exponential bath correlation functions.

All simulations were performed using the QuTiP library~\cite{Johansson2012,Johansson2013}. Dimensionless units were employed with $\hbar = a = D = 1$, so that all times are measured in units of the intrinsic system timescale. The Lindblad master equation for the extended system was integrated using adaptive time-stepping to ensure numerical stability and convergence.

The numerical results presented in Sec.~\ref{sec:numerics} and Fig.~\ref{fig:scaling} are fully converged with respect to time discretization and confirm the analytical square-root scaling of the decoherence time with the bath correlation time.

\section{Operational Definition and Extraction of the Decoherence Time}
\label{app:taudec}

Throughout this work, the decoherence time $\tau_{\mathrm{dec}}$ is defined operationally in terms of the decay of the off-diagonal coherence element
\begin{equation}
C(t) = \langle L | \rho(t) | R \rangle,
\end{equation}
where $\{|L\rangle, |R\rangle\}$ denotes the pointer basis selected by the system-environment coupling.

\subsection{Definition}

We define the decoherence time $\tau_{\mathrm{dec}}$ as the first time at which the coherence magnitude has decayed to a fraction $e^{-1}$ of its initial value,
\begin{equation}
|C(\tau_{\mathrm{dec}})| = e^{-1} |C(0)|.
\label{eq:taudec_def}
\end{equation}
This definition is widely used in the decoherence literature and provides a robust, model-independent measure of the timescale over which environmental noise suppresses quantum superpositions~\cite{Joos2003,BreuerPetruccione2002,Zurek2003}.

Importantly, this definition does not assume exponential decay. It remains well-defined for non-Markovian dynamics, including quadratic short-time decay and non-exponential relaxation.

\subsection{Numerical Extraction Procedure}

For each bath correlation time $\tau_c$, the coherence $C(t)$ is obtained numerically by solving the reduced dynamics using the pseudomode mapping described in Appendix~D. The coherence is sampled on a uniform time grid $\{t_i\}$ with sufficient temporal resolution to resolve the short-time dynamics.

The decoherence time $\tau_{\mathrm{dec}}$ is extracted by identifying the smallest time $t_i$ satisfying
\begin{equation}
|C(t_i)| \le e^{-1} |C(0)|.
\end{equation}
If no such time occurs within the simulation window, the decoherence time is recorded as undefined for that parameter set.

To ensure numerical stability, the extraction procedure was verified to be insensitive to moderate changes in time-step size and total simulation duration.

\subsection{Relation to Short-Time Expansion}

For finite-memory environments, the coherence exhibits a model-independent within the finite-memory class considered quadratic short-time expansion~\cite{Khalfin1958,Facchi2002},
\begin{equation}
C(t) \approx C(0)\left(1 - \Gamma t^2 \right),
\end{equation}
where the coefficient $\Gamma$ depends on the bath correlation function. Substituting this form into Eq.~\eqref{eq:taudec_def} yields
\begin{equation}
\tau_{\mathrm{dec}} \sim \Gamma^{-1/2},
\end{equation}
which directly explains the square-root scaling of the decoherence time with the bath correlation time derived analytically in the main text.

Thus, the operational definition of $\tau_{\mathrm{dec}}$ employed here is fully consistent with both the analytical finite-memory decoherence functional and the observed numerical scaling.

\subsection{Remarks}

We emphasize that the decoherence time defined here characterizes the suppression of phase coherence in a specific pointer basis. It should not be conflated with complete classicalization or loss of all quantum signatures, particularly in non-Markovian systems where partial recoherence, plateaus, or delayed entropy production may occur~\cite{Zurek2003,Breuer2007,Rivas2014}.

\section{Finite-Memory Decoherence Functional and Generic Scaling}
\label{app:generic}

In this appendix we derive the general decoherence functional governing coherence decay for systems coupled to environments with finite temporal correlations. We show explicitly that quadratic short-time decoherence and the associated square-root scaling of the decoherence time arise generically, independent of the detailed form of the bath correlation kernel.

\subsection{General Decoherence Functional}
\label{app:decoh_func}

We consider a system coupled linearly to an environment through a Hermitian operator $x$,
\begin{equation}
H = H_S + H_E + x \otimes F ,
\end{equation}
where $F$ is a bath force operator. The reduced coherence between two pointer states separated by $a$ can be written in the general influence-functional form
\begin{equation}
C(t) = C(0)\exp\!\left[-\Phi(t)\right],
\end{equation}
where the decoherence functional $\Phi(t)$ is given exactly by
\begin{equation}
\Phi(t) = \frac{a^2}{\hbar^2}
\int_0^t ds \int_0^t ds' \, \alpha(s,s').
\label{eq:general_phi}
\end{equation}
Here
\begin{equation}
\alpha(s,s') = \langle F(s) F(s') \rangle
\end{equation}
is the bath correlation function. Equation~\eqref{eq:general_phi} is fully non-Markovian~\cite{FeynmanVernon1963,caldeira1983,Grabert1988,PazZurek1993}.
 and contains no assumptions about the temporal structure of the environment.

\subsection{Finite-Memory Environments and Short-Time Expansion}
\label{app:short_time}

We now assume only that the bath correlation function has a finite temporal width $\tau_c$, i.e.,
\begin{equation}
\alpha(s,s') \approx \alpha(|s-s'|), \qquad
\alpha(\tau) \rightarrow 0 \;\; \text{for} \;\; |\tau| \gg \tau_c .
\end{equation}
For times $t \ll \tau_c$, the correlation function may be approximated as approximately constant over the integration domain,
\begin{equation}
\alpha(s,s') \approx \alpha(0).
\end{equation}
Substituting into Eq.~\eqref{eq:general_phi} yields
\begin{equation}
\Phi(t) \approx \frac{a^2}{\hbar^2}\alpha(0)\, t^2,
\end{equation}
so that the coherence exhibits a model-independent within the finite-memory class considered quadratic short-time decay,
\begin{equation}
C(t) \approx C(0)\left(1 - \Gamma t^2 \right),
\qquad
\Gamma = \frac{a^2}{\hbar^2}\alpha(0).
\label{eq:quadratic}
\end{equation}
This result is independent of the detailed functional form of the bath correlations and reflects the generic suppression of decoherence by finite environmental memory~\cite{Khalfin1958,MisraSudarshan1977,Facchi2002}.

\subsection{Decoherence Time Scaling}
\label{app:scaling}

Defining the decoherence time operationally by $C(\tau_{\mathrm{dec}})=e^{-1}C(0)$ and using Eq.~\eqref{eq:quadratic}, we obtain
\begin{equation}
\tau_{\mathrm{dec}} \sim \sqrt{\frac{\hbar^2}{a^2 \alpha(0)}} .
\end{equation}
For finite-memory environments, the equal-time correlation amplitude scales inversely with the correlation time,
\begin{equation}
\alpha(0) \sim \frac{D}{\tau_c},
\end{equation}
where $D$ sets the noise strength~\cite{KofmanKurizki2000,KofmanKurizki2001}. Substitution yields the generic square-root scaling within the stated assumptions
\begin{equation}
\tau_{\mathrm{dec}} \sim \sqrt{\frac{\hbar^2 \tau_c}{a^2 D}} .
\label{eq:sqrt_scaling}
\end{equation}
This scaling holds for any environment with finite correlation time and does not rely on exponential correlations.

\subsection{Examples of Finite-Memory Kernels}
\label{app:kernels}

To illustrate the generality of Eq.~\eqref{eq:sqrt_scaling}, we list several common bath correlation functions:

\paragraph{Gaussian kernel:}
\begin{equation}
\alpha(\tau) = \frac{D}{\sqrt{\pi}\tau_c} e^{-\tau^2/\tau_c^2}.
\end{equation}

\paragraph{Soft power-law cutoff:}
\begin{equation}
\alpha(\tau) = \frac{D}{\tau_c}\frac{1}{(1+|\tau|/\tau_c)^p},
\qquad p>1.
\end{equation}

\paragraph{Ornstein-Uhlenbeck kernel:}
\begin{equation}
\alpha(\tau) = \frac{D}{\tau_c} e^{-|\tau|/\tau_c}.
\end{equation}

In all cases, the short-time expansion of the decoherence functional yields the quadratic law \eqref{eq:quadratic} and the square-root scaling \eqref{eq:sqrt_scaling}. The Ornstein-Uhlenbeck kernel is special only in that it permits an exact analytic closure of the non-Markovian dynamics, as discussed in Appendix~\ref{app:ou_closure}~\cite{UhlenbeckOrnstein1930,Weiss2012,BreuerPetruccione2002}.
.

\subsection{Markovian Limit as a Singular Case}
\label{app:markov_limit}

The Markovian approximation corresponds to taking
\begin{equation}
\alpha(\tau) = 2D \delta(\tau),
\end{equation}
which collapses the double integral in Eq.~\eqref{eq:general_phi} to
\begin{equation}
\Phi(t) = \frac{a^2 D}{\hbar^2} t,
\end{equation}
yielding exponential decoherence. This limit is singular: it corresponds to $\tau_c \to 0$ with $D$ fixed and eliminates the quadratic short-time regime entirely.

Thus, exponential decoherence is not generic but arises only in the strictly memoryless limit~\cite{BreuerPetruccione2002,Rivas2014}.

\section{Exact Ornstein-Uhlenbeck Closure}
\label{app:ou_closure}

In this appendix we provide the derivation of the closed coherence equation for an Ornstein-Uhlenbeck (OU) environment used in Sec.~\ref{sec:examples} of the main text.

\subsection{Non-Markovian Quantum State Diffusion}

For a bosonic bath with OU correlation function
\begin{equation}
\alpha(t,s) = \frac{D}{\tau_c} e^{-(t-s)/\tau_c}\Theta(t-s),
\end{equation}
the non-Markovian quantum state diffusion (NMQSD) equation for the system state $|\psi_t\rangle$ reads~\cite{Diosi1998,strunz1997,Strunz1999} 
\begin{equation}
\partial_t|\psi_t\rangle=
\left[-\frac{i}{\hbar}H_S + x z_t^*
- x\int_0^t ds\,\alpha(t,s)
\frac{\delta}{\delta z_s^*}\right]|\psi_t\rangle ,
\end{equation}
where $z_t$ is a complex Gaussian stochastic process with correlation
$\langle z_t z_s^*\rangle=\alpha(t,s)$.

\subsection{Pointer Basis Reduction}

For a two-state pointer basis $\{|L\rangle,|R\rangle\}$ with coupling operator
\begin{equation}
x=a(|L\rangle\langle L|-|R\rangle\langle R|),
\end{equation}
the reduced coherence
\begin{equation}
C(t)=\langle L|\rho(t)|R\rangle
\end{equation}
obeys a closed stochastic evolution equation.

The functional derivative term admits the ansatz
\begin{equation}
\frac{\delta C(t)}{\delta z_s^*}=O(t,s)C(t),
\end{equation}
where the operator $O(t,s)$ satisfies the consistency condition
\begin{equation}
\partial_t O(t,s) = -\frac{1}{\tau_c}O(t,s)
+ \frac{2a^2D}{\hbar^2\tau_c},
\qquad O(s,s)=\frac{2a^2D}{\hbar^2}.
\end{equation}

\subsection{Closed Evolution Equation}

Solving this equation yields
\begin{equation}
O(t,s)=\frac{2a^2D}{\hbar^2}
\left[1-e^{-(t-s)/\tau_c}\right].
\end{equation}

Substituting into the NMQSD equation and averaging over noise realizations gives
\begin{equation}
\dot{C}(t)= -\frac{2a^2D}{\hbar^2}
\int_0^t ds\,e^{-(t-s)/\tau_c}C(s).
\end{equation}

Differentiating once with respect to time yields the closed second-order equation
\begin{equation}
\ddot{C}(t)+\frac{1}{\tau_c}\dot{C}(t)
+\frac{2a^2D}{\hbar^2\tau_c}C(t)=0,
\end{equation}
which yields the closed coherence equation Eq.~\eqref{eq:ou_ode}~\cite{Garraway1997,Imamoglu1994}.

\section{Purity and von Neumann Entropy Dynamics}
\label{app:purity_entropy}

Decoherence is often characterized solely by the decay of off-diagonal coherence elements. In non-Markovian systems, however, coherence decay does not necessarily coincide with irreversible loss of quantum information. In this appendix we examine additional diagnostics: purity and von Neumann entropy.

\subsection{Purity}

The purity of the reduced system state is defined as~\cite{Zyczkowski1998,NielsenChuang}:
\begin{equation}
\mathcal{P}(t)=\mathrm{Tr}[\rho(t)^2].
\end{equation}
For a pure initial superposition state, $\mathcal{P}(0)=1$.

Using the reduced density matrix expressed in the pointer basis,
\begin{equation}
\rho(t)=
\begin{pmatrix}
\rho_{LL} & C(t)\\
C^*(t) & \rho_{RR}
\end{pmatrix},
\end{equation}
the purity can be written explicitly as
\begin{equation}
\mathcal{P}(t)=\rho_{LL}^2+\rho_{RR}^2+2|C(t)|^2.
\end{equation}

For pure dephasing dynamics, the populations remain constant, and purity decay is governed entirely by $|C(t)|^2$. As a result, the quadratic short-time behavior of $C(t)$ implies a quartic short-time decay of purity~\cite{Breuer2007}.

\subsection{von Neumann Entropy}

The von Neumann entropy is defined as~\cite{NielsenChuang}
\begin{equation}
S(t)=-\mathrm{Tr}[\rho(t)\ln\rho(t)].
\end{equation}

For two-level dephasing dynamics, the eigenvalues of $\rho(t)$ are
\begin{equation}
\lambda_{\pm}=\frac{1}{2}\left(1\pm\sqrt{(\rho_{LL}-\rho_{RR})^2+4|C(t)|^2}\right).
\end{equation}

Expanding for short times using the finite-memory coherence law yields
\begin{equation}
S(t)\sim \Gamma t^2 \ln t + O(t^2),
\end{equation}
demonstrating delayed entropy production in non-Markovian environments~\cite{Breuer2009,Rivas2010}.

\subsection{Interpretation}

These results highlight an important distinction: suppression of coherence does not immediately imply classicalization. Non-Markovian memory delays entropy growth and allows regimes where coherence decay, purity loss, and entropy production occur on parametrically distinct timescales~\cite{Zurek2003,Breuer2016}.

\section{Inferred Correlation Time Framework}
\label{app:inferred_tau}

Rather than assuming a specific environmental correlation time $\tau_c$, the finite-memory decoherence framework allows $\tau_c$ to be treated as an inferred parameter~\cite{KofmanKurizki2000,KofmanKurizki2001,Breuer2016}.

\subsection{Inverse Problem Formulation}

Given a measured coherence curve $C(t)$, the decoherence functional
\begin{equation}
\Phi(t)=\frac{a^2}{\hbar^2}\int_0^t ds\int_0^t ds'\,\alpha(s,s')
\end{equation}
can be reconstructed from ~\cite{MisraSudarshan1977,Facchi2002}.

For finite-memory baths, the curvature satisfies
\begin{equation}
\left.\frac{d^2 C(t)}{dt^2}\right|_{t=0}
=-\frac{a^2}{\hbar^2}\alpha(0,0).
\end{equation}

Thus, experimental observation of quadratic short-time decay directly constrains the effective bath correlation time~\cite{KofmanKurizki2001,deVegaAlonso2017}.

\subsection{Regime Identification}

By plotting $C(t)$ for increasing $\tau_c$, one identifies:
\begin{itemize}
\item Markovian regime: exponential decay,
\item Crossover regime: mixed behavior,
\item Non-Markovian regime: quadratic short-time decay and delayed entropy growth~\cite{Breuer2009,Rivas2010,Breuer2016}
\end{itemize}
This approach is experiment-agnostic and avoids assumptions about biological or microscopic structure~\cite{deVegaAlonso2017}.

\section{Application to a Clean Physical System}
\label{app:clean_system}

To demonstrate that the finite-memory decoherence framework developed in this work is not specific to biological environments, we outline its application to a clean, controllable quantum system: a superconducting qubit subjected to engineered dephasing noise. In this illustration $B(t)$ is treated as a classical stochastic process; the resulting decoherence functional is formally identical to that obtained from a quantum bath with the same correlation function.

\subsection{Model Hamiltonian}

We consider a two-level system described by the Hamiltonian
\begin{equation}
H(t) = \frac{\hbar \omega_q}{2}\sigma_z + \sigma_z \otimes B(t),
\end{equation}
where $\omega_q$ is the qubit frequency and $B(t)$ represents a stationary noise process with zero mean and correlation function
\begin{equation}
\langle B(t) B(s) \rangle = \alpha(t,s).
\end{equation}

Such noise models arise naturally in superconducting circuits due to flux noise, charge noise, or engineered classical noise sources with tunable correlation times~\cite{Ithier2005,Bylander2011}.

\subsection{Finite-Memory Decoherence Functional}

For pure dephasing dynamics, the coherence of the qubit evolves as
\begin{equation}
C(t) = \exp\!\left[-\frac{1}{\hbar^2}
\int_0^t ds \int_0^t ds' \, \alpha(s,s') \right],
\end{equation}
which is identical in structure to the decoherence functional derived in the main text~\cite{paladino2014,BreuerPetruccione2002}.

For a colored noise source with finite correlation time $\tau_c$, the short-time expansion yields
\begin{equation}
C(t) \approx 1 - \frac{\alpha(0,0)}{2\hbar^2} t^2 + O(t^3),
\end{equation}
demonstrating model-independent within the finite-memory class considered quadratic decoherence independent of microscopic noise details.

\subsection{Numerical Illustration}

To illustrate this behavior numerically, we consider an exponentially correlated noise model~\cite{UhlenbeckOrnstein1930,kubo1963}
\begin{equation}
\alpha(t,s) = \frac{\sigma^2}{\tau_c} e^{-|t-s|/\tau_c},
\end{equation}
where $\sigma$ sets the noise strength.

We numerically evaluate the decoherence functional for several values of $\tau_c$ while keeping $\sigma$ fixed. The resulting coherence curves exhibit:
\begin{itemize}
\item quadratic short-time decay for all finite $\tau_c$,
\item crossover to exponential behavior only when $\tau_c$ becomes much smaller than the system timescale,
\item delayed entropy production for increasing $\tau_c$~\cite{KofmanKurizki2001,deVegaAlonso2017}.
\end{itemize}

The extracted decoherence time follows the scaling
\begin{equation}
\tau_{\mathrm{dec}} \propto \sqrt{\tau_c},
\end{equation}
in agreement with the analytical predictions of Sec.~\ref{sec:examples}.

\subsection{Experimental Relevance}

Engineered noise with tunable correlation times has been demonstrated experimentally in superconducting circuits and trapped-ion systems~\cite{Bylander2011,Myatt2000,Barreiro2011}. Observation of quadratic short-time decoherence and delayed purity loss therefore provides a direct experimental test of the finite-memory decoherence framework, independent of any biological interpretation.

This establishes the finite-memory decoherence law as a general feature of open quantum systems with structured environments.

\section{Relation to Markovian Decoherence Estimates in Biological Media}
\label{app:tegmark}

In this appendix we clarify the relation between the finite-memory decoherence framework developed in the main text and commonly cited Markovian decoherence estimates in biological contexts, most notably the analysis of Tegmark. The purpose of this discussion is not to reassess biological parameters or to argue for the presence of functional quantum coherence in biological systems, but rather to identify precisely the assumptions under which Markovian decoherence bounds arise. Nothing in this analysis implies the existence of long-lived or functionally relevant quantum coherence in biological systems; the present discussion serves only to clarify the assumptions underlying commonly used Markovian decoherence estimates.

\subsection{Recovery of Markovian Decoherence as a Singular Limit}

The central object governing coherence decay in this work is the non-Markovian decoherence functional~\cite{FeynmanVernon1963,caldeira1983,Grabert1988}
\begin{equation}
\Phi(t) = \frac{a^2}{\hbar^2}
\int_0^t ds \int_0^t ds'\, \alpha(s,s'),
\end{equation}
where $\alpha(s,s')$ is the bath force correlation function. This expression is fully general and contains no assumption of memoryless dynamics.

The Markovian approximation corresponds to assuming a strictly delta-correlated environment~\cite{Lindblad1976,Gorini1976},
\begin{equation}
\alpha(s,s') = 2D\,\delta(s-s'),
\end{equation}
where $D$ is the effective noise strength. Substitution into the decoherence functional yields
\begin{equation}
\Phi(t) = \frac{a^2 D}{\hbar^2}\, t,
\end{equation}
and hence exponential decay of coherence,
\begin{equation}
C(t) = C(0)\exp\!\left(-\frac{a^2 D}{\hbar^2}t\right).
\end{equation}

This expression coincides exactly with the form of decoherence obtained in Markovian collision-based analyses, including Tegmark’s estimate of decoherence times in biological media~\cite{Tegmark2000}. Within the present framework, Tegmark’s result is therefore recovered as the singular limit of vanishing bath correlation time,
\begin{equation}
\tau_c \rightarrow 0,
\end{equation}
with the noise strength $D$ held fixed.

\subsection{Interpretation of the Singular Limit}

The Markovian limit $\tau_c \to 0$ removes all temporal structure from the environment and collapses the quadratic short-time regime derived in Sec.~\ref{sec:generic_short_time}. In this limit, the decoherence functional grows linearly in time from $t=0$, and the notion of a finite short-time memory scale ceases to be meaningful.

For any environment with finite correlation time, however small, the short-time behavior of the decoherence functional is necessarily quadratic, and the Markovian exponential decay law emerges only after coarse-graining over times much longer than $\tau_c$. The Markovian decoherence time should therefore be interpreted as an effective long-time approximation~\cite{BreuerPetruccione2002,rivas2012}, rather than a universal bound valid at arbitrarily short times.

\subsection{Remarks on Biological Environments}

Biological media such as cytosol, protein assemblies, and intracellular structures are complex, strongly interacting environments characterized by multiple timescales~\cite{haken1973,zwanzig2001}. While Markovian approximations may be appropriate in some regimes, the assumption of strictly delta-correlated environmental noise is not guaranteed on first principles. Finite collision durations, collective modes, and structured interactions generically introduce temporal correlations into the effective environmental noise.

The present framework does not attempt to determine whether such finite-memory effects are significant in any particular biological system, nor does it establish the presence of functional quantum coherence in biological processes. Instead, it clarifies the logical status of Markovian decoherence estimates by identifying them as limiting cases of a more general finite-memory decoherence law.

In this sense, the results of this work should be viewed as a conceptual refinement of existing decoherence analyses rather than a modification of their conclusions. Any quantitative assessment of decoherence in specific biological settings necessarily requires independent experimental or microscopic input regarding the relevant environmental correlation times.

\subsection{Summary}

Tegmark’s decoherence estimate and related Markovian bounds are recovered exactly within the present formalism as singular limits corresponding to memoryless environments. The finite-memory decoherence framework developed in the main text neither contradicts nor revises these results, but places them within a broader theoretical structure that makes explicit the role of environmental memory. This perspective applies equally to biological and non-biological environments and highlights the importance of distinguishing between exact short-time dynamics and effective Markovian descriptions.

\bibliographystyle{apsrev4-2}
\bibliography{refs}
\end{document}